\documentclass[12pt]{amsart}
\usepackage{amsmath}
\usepackage{graphicx}
\usepackage{tikz}
\begin{document}

\title[Integrable Chiral Potts Model]{About 30 Years of Integrable Chiral Potts Model,
Quantum Groups at Roots of Unity and Cyclic Hypergeometric Functions}

\author{Helen Au-Yang and Jacques H H Perk}

\address{Department of Physics, Oklahoma State University,
145 Physical Sciences, Stillwater, OK 74078-3072, USA}

\email{perk@okstate.edu helenperk@yahoo.com}

\thanks{This report summarizes our two contributions to
the International Conference on Subfactor Theory in Mathematics and Physics,
July 13-17, 2015, Qinhuangdao, China, honoring Dr.\ Vaughan F.R.~ Jones.
We are most grateful to Dr.\ Zheng-Wei Liu for all his help and effort that made
this meeting such a success.}

\begin{abstract}
In this paper we discuss the integrable chiral Potts model, as it clearly relates
to how we got befriended with Vaughan Jones, whose birthday we celebrated
at the Qinhuangdao meeting. Remarkably we can also celebrate the birthday
of the model, as it has been introduced about 30 years ago as the first solution
of the star-triangle equations parametrized in terms of higher genus functions.
After introducing the most general checkerboard Yang--Baxter equation,
we specialize to the star-triangle equation, also discussing its relation with
knot theory. Then we show how the integrable chiral Potts model leads
to special identities for basic hypergeometric series in the $q$ a root-of-unity limit.
Many of the well-known summation formulae for basic hypergeometric series do
not work in this case. However, if we require the summand to be periodic,
then there are many summable series. For example, the
integrability condition, namely, the star-triangle equation, is a summation
formula for a well-balanced ${}_4\Phi_3$ series. We finish with a few remarks
about the relation with quantum groups at roots of unity.

\end{abstract}

\maketitle

\def\be{\begin{equation}}
\def\ee{\end{equation}}
\def\ba{\begin{eqnarray}}
\def\ea{\end{eqnarray}}

\def\hidefrac#1#2{\begin{array}{@{}c@{}}#1\\#2\end{array}}
\font\my=cmr12 at 14pt
\def\myphi{\hbox{\my\char'010}}
\def\hypp#1#2#3#4#5{{}_{#1}\myphi_{#2}\!\!\left[{\hidefrac{#3}{#4}};#5\right]}
\def\hypg#1#2#3{{}_{p+1}\myphi_p\left[{\hidefrac{#1}{#2}};#3\right]}
\def\hypf#1#2#3#4#5{{}_{#1}F_{#2}\!\!\left[{\hidefrac{#3}{#4}};#5\right]}

\section{How we got to know Vaughan Jones}

In 1988 one of us found a preprint by Vaughan Jones, ``On a Certain Value of the
Kauffman Polynomial'' \cite{Jones1}. We immediately saw that the metaplectic
representation for $p=5$ therein had to be related to our chiral Potts model work \cite{AMPTY,BPA} and it soon became clear that it was even related to the 5-state
Fateev--Zamolodchikov model \cite{FZ}.

After reporting this to Vaughan we received an invitation to the Workshop on
Integrable Systems in Statistical Mechanics, Quantum Field Theory, and Knot Theory,
at MSRI, UC Berkeley, January 1989. There we both had detailed further
discussions on the relationships between knot theory and integrable models of statistical
mechanics and some of that got incorporated in the paper
``On Knot Invariants Related to Some Statistical Mechanical Models'' \cite{Jones2}.

One evening during the workshop, Vaughan invited us both to his apartment,
but Helen could not go as our baby was not well. At one point Vaughan came
up with five copies of H\"andel's Messiah for five of us present. He then
assigned the four voice parts, ``Jacques, do you want to sing tenor or bass?''\
and Anthony Wasserman got the piano accompaniment. Thus we performed
the entire piece from cover to cover. It was a deeply spiritual experience
with the entire text taken from the King James Version of the Bible.

We have met Vaughan later at several other meetings and have played several
games of snooker together. Once during a conference honoring Baxter
in Canberra we had a bus trip to Tidbinbilla. Vaughan said, ``Helen, you have
to sit somewhere else; I have to sit next to Jacques."\ and we sang several
parts of the Messiah together. Even during the Great Wall expedition of the
Qinhuangdao conference we sang `Comfort ye' and `Every valley' in the bus after Helen
fell asleep and her earplug came loose from her iPad.

\section{Yang--Baxter integrable statistical mechanics models}

\begin{figure}
\begin{center}
\includegraphics[ width=4in]{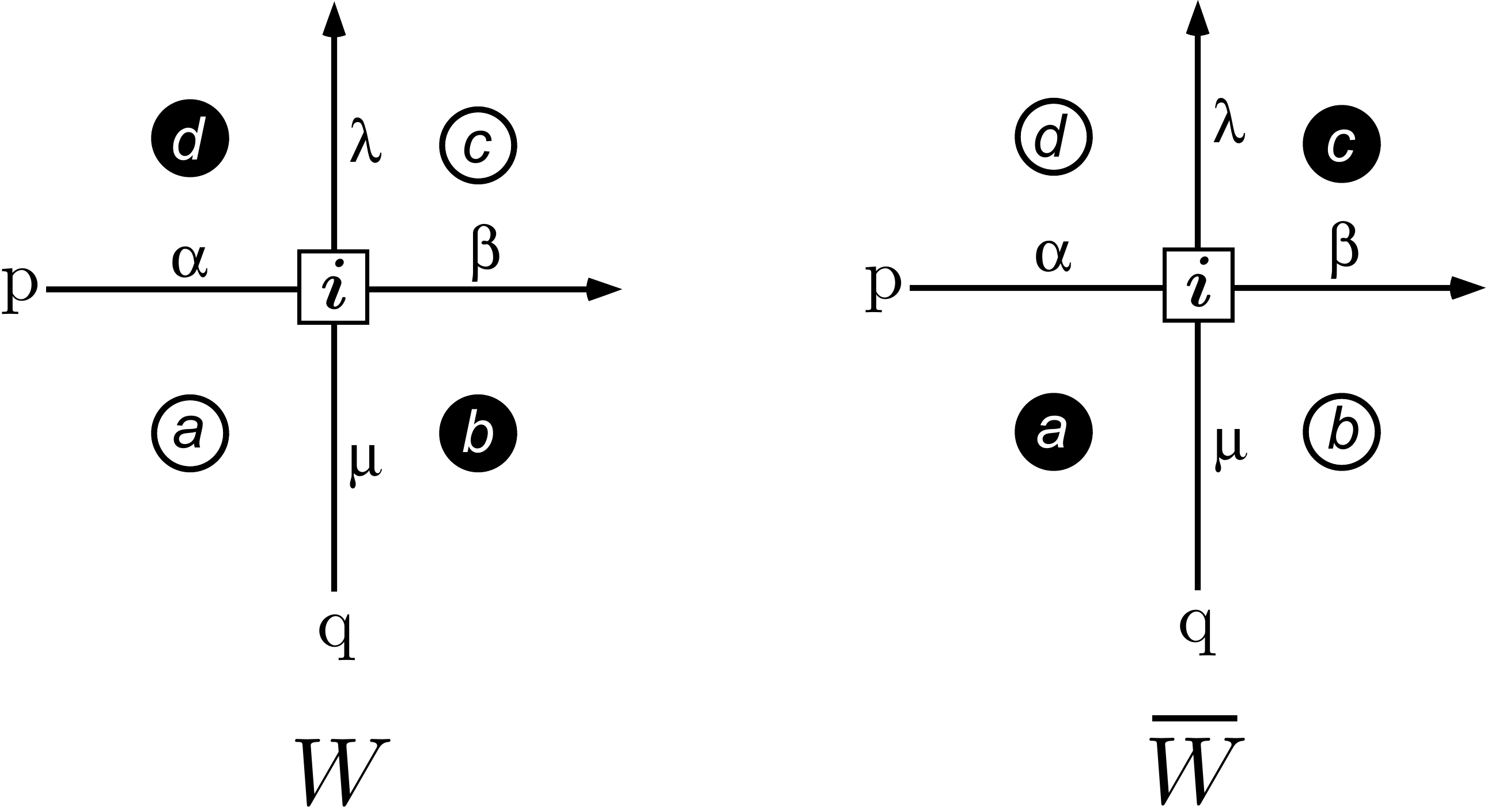}
\end{center}
\caption{The Boltzmann weights $W{}^{(i)}{}^{\lambda\beta}_{\alpha\mu}|{}^{dc}_{ab}(p,q)$
and $\overline W{}^{(i)}{}^{\lambda\beta}_{\alpha\mu}|{}^{dc}_{ab}(p,q)$}
associated with the crossing of oriented rapidity lines with rapidities p and q.
Spin variables live on line pieces, faces and vertices. Faces are colored
alternatingly black and white in a checkerboard pattern.
\label{fig1}\end{figure}
The Yang--Baxter equation \cite{Baxterbook,PA} is a generalization of Artin's braid
equation in knot theory with spectral variables called rapidities $\mathrm{p},
\mathrm{q},\cdots$, living on oriented lines, see figure \ref{fig1}. To each crossing
of the rapidity lines one assigns Boltzmann weights
$W{}^{(i)}{}^{\lambda\beta}_{\alpha\mu}|{}^{dc}_{ab}(p,q)$
or $\overline W{}^{(i)}{}^{\lambda\beta}_{\alpha\mu}|{}^{dc}_{ab}(p,q)$
depending on a black-white checkerboard coloring assigned to the faces.
Most generally, one can have spin variables on all faces, vertices and line pieces,
while each spin variable can be chosen independently from a finite or
infinite set.

\begin{figure}
\begin{center}
\includegraphics[ width=4in]{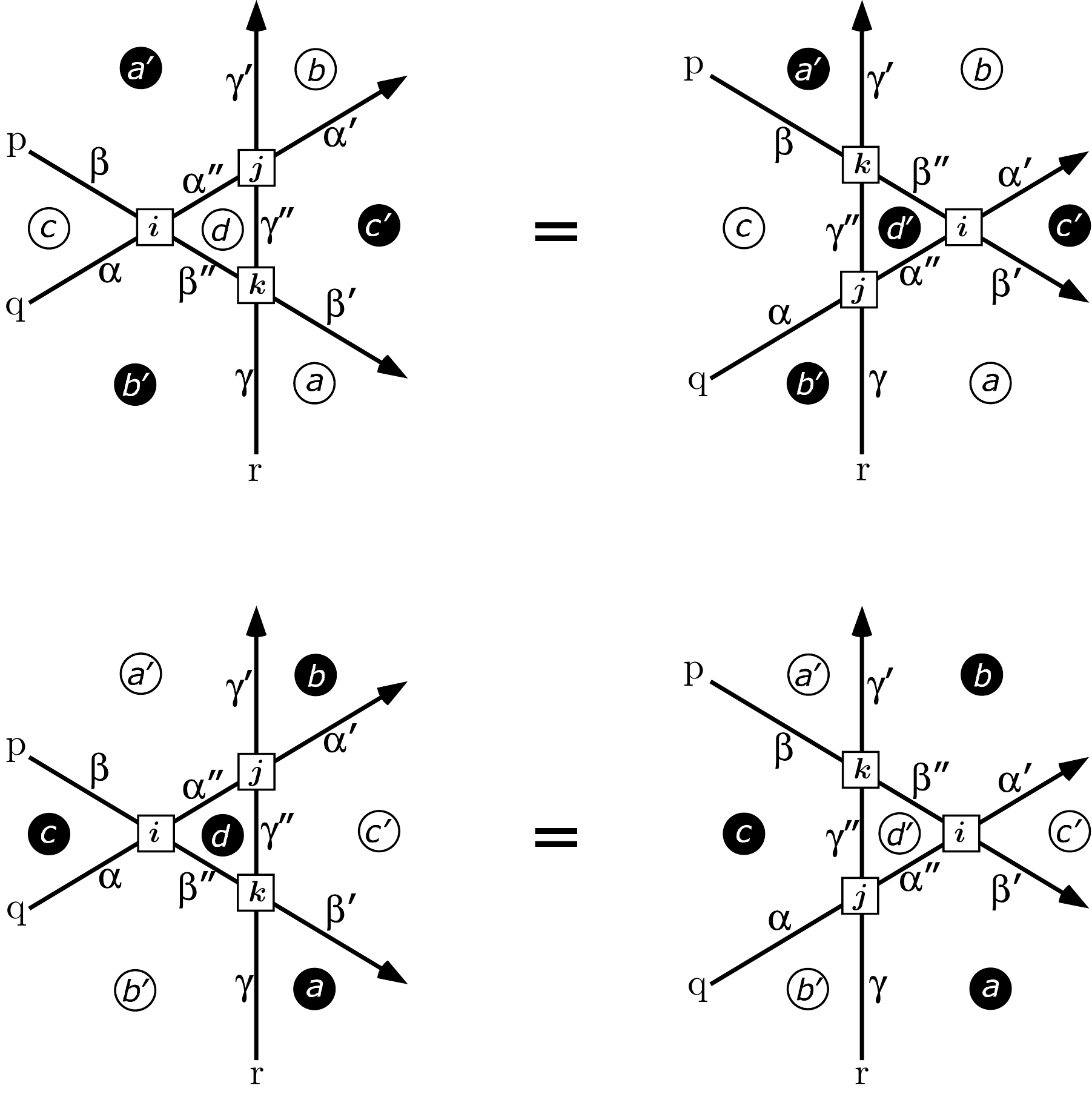}
\end{center}
\caption{Pictorial representation of the checkerboard Yang--Baxter equations.
State sums over spins on the vertices and on internal faces and line pieces are
assumed.}
\label{fig2}\end{figure}

One can sum over the spin $i$ at the vertex, obtaining the IRF-vertex model
with new Boltzmann weights
$\sum_iW{}^{(i)}{}^{\lambda\beta}_{\alpha\mu}|{}^{dc}_{ab}(p,q)\equiv
W{}^{\lambda\beta}_{\alpha\mu}|{}^{dc}_{ab}(p,q)$,
$\sum_i\overline W{}^{(i)}{}^{\lambda\beta}_{\alpha\mu}|{}^{dc}_{ab}(p,q)\equiv
\overline W{}^{\lambda\beta}_{\alpha\mu}|{}^{dc}_{ab}(p,q)$.
Such a reduction can always be applied and has been useful in certain
spin models, see e.g.\ \cite{BaxFF,BBP,JPBax}.

Assuming that there are no spins at the intersections of rapidity lines, one
can define several special cases \cite{PA}. First, if the spins on the line pieces
($\alpha,\beta,\cdots$) only take a single value, one can omit them,
arriving at the checkerboard Interaction-Round-a-Face (IRF) model
with weights $W{}^{dc}_{ab}(p,q)$, $\overline W{}^{dc}_{ab}(p,q)$.
Second, assuming all spins on the faces have a single value (and thus
can be omitted) one receives the checkerboard vertex model with weights
$W{}^{\lambda\beta}_{\alpha\mu}(p,q)$, 
$\overline W{}^{\lambda\beta}_{\alpha\mu}(p,q)$. Forgetting about the coloring
of the faces one gets the usual vertex model with
$W{}^{\lambda\beta}_{\alpha\mu}(p,q)=
\overline W{}^{\lambda\beta}_{\alpha\mu}(p,q)$.  Finally, if one only leaves
spins on the black (or white) faces, one ends up with a spin model.

The most general Yang--Baxter equation is depicted in figure \ref{fig2} and can
be expressed more complicatedly in formula as

\ba
&&\sum_i\sum_j\sum_k\sum_{\alpha''}\sum_{\beta''}\sum_{\gamma''}\sum_{d}
W{}^{(i)}{}^{\alpha''\!\beta''}_{\beta\,\,\alpha}\!|{}^{a'd}_{c\,b'}(p,q)
\label{eq1}\\&&\hspace{6em}\times\,
W{}^{(j)}{}^{\gamma'\,\alpha'}_{\alpha''\gamma''}\!|{}^{a'b}_{d\,c'}(q,r)
\,\overline W{}^{k)}{}^{\gamma''\beta'}_{\beta''\gamma}\!|{}^{d\,c'}_{b'a}(p,r)
\nonumber\\
&&\quad=R(p,q,r)\,
\sum_i\sum_j\sum_k\sum_{\alpha''}\sum_{\beta''}\sum_{\gamma''}\sum_{d'}
\overline W{}^{(i)}{}^{\alpha'\,\beta'}_{\beta''\alpha''}\!|{}^{b\,c'}_{d'a}(p,q)
\nonumber\\&&\hspace{6em}\times\,
\overline W{}^{(j)}{}^{\gamma''\!\alpha''}_{\alpha\,\,\gamma}\!|{}^{c\,d'}_{b'a}(q,r)
\,W{}^{(k)}{}^{\gamma'\beta''}_{\beta\,\gamma''}\!|{}^{a'b}_{c\,d'}(p,r),
\nonumber\\
\cr&&\overline R(p,q,r)\,
\sum_i\sum_j\sum_k\sum_{\alpha''}\sum_{\beta''}\sum_{\gamma''}\sum_{d}
\overline W{}^{(i)}{}^{\alpha''\!\beta''}_{\beta\,\,\alpha}\!|{}^{a'd}_{c\,b'}(p,q)
\label{eq2}\\&&\hspace{6em}\times\,
\overline W{}^{(j)}{}^{\gamma'\,\alpha'}_{\alpha''\gamma''}\!|{}^{a'b}_{d\,c'}(q,r)
\,W{}^{(k)}{}^{\gamma''\beta'}_{\beta''\gamma}\!|^{d\,c'}_{b'a}(p,r)\nonumber\\
&&\quad=\sum_i\sum_j\sum_k\sum_{\alpha''}\sum_{\beta''}\sum_{\gamma''}\sum_{d'}
W{}^{(i)}{}^{\alpha'\,\beta'}_{\beta''\alpha''}\!|{}^{b\,c'}_{d'a}(p,q)
\cr&&\hspace{6em}\times\,
W{}^{(j)}{}^{\gamma''\!\alpha''}_{\alpha\,\,\gamma}\!|{}^{c\,d'}_{b'a}(q,r)
\,\overline W{}^{(k)}{}^{\gamma'\beta''}_{\beta\,\gamma''}\!|{}^{a'b}_{c\,d'}(p,r).
\nonumber
\ea

One can read more about this and the various formal equivalences and specializations
in \cite{PA,AP}. As said before, the sums over $i$, $j$ and $k$ can be taken introducing
new sum weights and one can specialize further to get the various special versions of the
Yang--Baxter equation in the literature.

Finally, in some examples that we have studied the scalar factors $R(p,q,r)$ and
$\overline R(p,q,r)$ can be factorized and absorbed into the $W$'s
and $\overline W$'s by properly redefining these. Also (\ref{eq1}) and (\ref{eq2})
are the same equation in the integrable chiral Potts model, for example.

\section{Integrable chiral Potts model}

Let us from now on specialize to the integrable chiral Potts model
\cite{AMPTY,BPA}. Potts means here that there is a translation invariance
in the spin variables, meaning that each weight is of the form
$W_a^b(p,q)\equiv W_{pq}(a-b)$ or
$\overline W_a^b(p,q)\equiv\overline W_{pq}(a-b)$, depending only on the difference
of two spin variables $a$ and $b$ modulo an integer $N$ and integrability
implies the existence of the two rapidities $p$ and $q$. The chiral property means
that there is no reflection invariance, i.e.\ $W_{pq}(a-b)\ne W_{pq}(b-a)$,
$\overline W_{pq}(a-b)\ne\overline W_{pq}(b-a)$
in general.

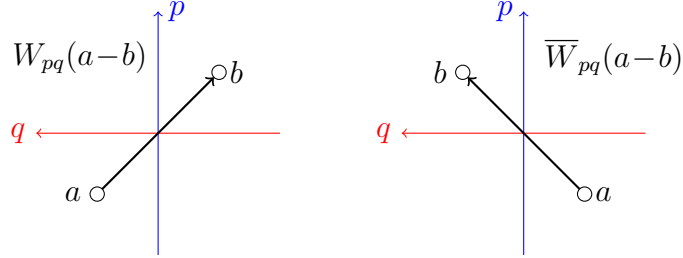
\begin{figure}
\begin{center}\hspace*{-0.2in}
\begin{tikzpicture}[scale=1.35]
\draw  [->,red] (3.6,0) -- (1.2,0) node[left]  {$q$} ;
 \draw [->, red] (0,0) -- (-2.4,0)  node[left]  {$q$};
\draw [->, blue] (2.4,-1.2) -- (2.4,1.2) node[left] {$p$} ;
 \draw [->, blue] (-1.2,-1.2) -- (-1.2,1.2) node[right] {$p$} ;
 
\node[left] at (-1.2,0.75) {$ W_{pq}(a\!-\!b)$};
\node[right] at (2.5,0.75) {$ \overline W_{pq}(a\!-\!b)$};
\draw (1.8,0.6) circle (2pt);\node [left]at (1.75,0.6) {$b$};
\draw(3,-0.6) circle (2pt);\node [right]at (3,-0.6) {$a$};
\draw (-1.8,-0.6) circle (2pt);\node [left]at  (-1.85,-0.6) {$a$};
\draw (-0.6,0.6) circle (2pt);\node [right]at  (-0.6,0.6) {$b$};

\draw [->,thick] (-1.75,-0.55) -- (-0.65,0.55)  ;
\draw [->,thick] (2.95,-0.55) -- (1.85,0.55)  ;
\end{tikzpicture}
\end{center}
\caption{The chiral Potts model weights.}
\label{fig3}\end{figure}

The two weights are depicted in figure \ref{fig3} and have the form
\ba
&&\displaystyle W_{pq}(n)=W_{pq}(0)\prod^{n}_{j=1}\biggl(\frac{\mu_p}{\mu_q}\cdot
\frac{y_q-x_p\omega^j}{y_p-x_q\omega^j}\biggr),
\label{Bweights}\\
&&\displaystyle {\overline W}_{pq}(n)={\overline W}_{pq}(0)\prod^{n}_{j=1}\biggl(\mu_p\mu_q\cdot
\frac{\omega x_p-x_q\omega^j}{y_q-y_p\omega^j}\biggr),\nonumber
\ea
where $n=a-b$, $\omega=\exp(2\pi\sqrt{-1}/N)$ is an $N$th root of unity,
and the two rapidities $p=(x_p,y_p,\mu_p)$ and $q=(x_q,y_q,\mu_q)$
lie on the high-genus curve
\be
x_p^N+y_p^N=k(1+x_p^Ny_p^N),\qquad\displaystyle
\mu_p^N=\frac{k'}{1-k\,x_p^N}=\frac{1-k\,y_p^N}{k'},
\ee
with some $k$ and $k'$ satisfying $k^2+k'^2=1$. Here we shall choose
the normalization $W_{pq}(0)=\overline W_{pq}(0)=1$.

These weights satisfy the Reflection Relation
\be
W_{pp}(a-b)=1,\quad
W_{pq}(a-b)W_{qp}(a-b)=1,
\ee
and Inversion Relation
\be
\sum_{b=0}^{N-1}\overline W_{pq}(a-b)\overline W_{qp}(b-c)=r_{pq}\delta_{a,c}.
\ee
These relations correspond to Reidemeister moves I and II of knot theory. Indeed,
\begin{figure}
\begin{center}
 \begin{tikzpicture}[scale=0.7]
\draw  [->,blue,thick](0,-3) -- (0,0).. controls +(up:1mm) and +(left:5mm).. (0.5,0.5);
\draw  [->,blue,thick] (0.5,0.5).. controls ++(right:1mm) and ++(up:5mm)..(1,0)
.. controls ++(down:5mm) and ++(down:5mm) .. (0,0)-- (0,3)  node[above] {$p$} ;
 \node[below] at (2,0.5) {$=$} ;
 \draw [->,blue,thick] (4,-3) -- (4,3) node[above] {$p$};
 \filldraw[green] (0.4,0) circle (2pt) ;\node[right] at (0.27,0.1) {$a$} ; 
\filldraw[green] (-0.6,0) circle (2pt) ;\node[left] at (-0.5,0.1) {$b$};
 \draw [->,thin,gray] (-0.6,0) -- (0.4,0);
\node[below] at (2,0.5) {$=$} ;
 \filldraw[green] (3.6,0) circle (2pt) ;\node[left] at (3.6,0) {$b$};
 \filldraw (4.4,0) circle (2pt) ;\node[right] at (4.4,0) {$c$};
  \draw [<-, thin, gray] (1,1.2) ..controls ++(left:4mm) and ++(up:5mm).. (0,0)
  .. controls+(down:5mm) and +(left:4mm) .. (1,-1.2);
   \filldraw (1,1.2) circle (2pt) ;\node[above] at (1,1.2) {$c$} ; 
   \filldraw (1,-1.2) circle (2pt);\node[below] at (1,-1.2) {$c'$} ;
\end{tikzpicture}
\hspace{1.7cm}
\begin{tikzpicture}[scale=0.7]
\draw  [->,blue,thick](3.5,-3)  .. controls (2,0).. (3.5,3) node[right]  {$p$} ;
\draw  [->,red,thick] (2,-3).. controls  (3.5,0)..(2,3) node[left] at (2,3) {$q$} ;
 \filldraw [green](1.9,0) circle (2pt) ;\node[above] at (1.78,0) {$b$};
 \filldraw  (2.75,0) circle (2pt) ;\node[above] at (2.8,0) {$d$};
   \filldraw [green] (3.6,0) circle (2pt) ;\node[above] at (3.72,0) {$c$};
    \filldraw (2.75,-2.6) circle (2pt) ;\node[below] at (2.8,-2.6) {$a$};
     \filldraw (2.75,2.6) circle (2pt) ;\node[above] at (2.8,2.6) {$a'$};
\draw  [->,thin,gray](1.9,0)..controls ++(up:4mm) and ++(left:5mm).. (2.75,1.45) 
..controls ++(right:4mm) and ++(up:5mm)..(3.6,0);
\draw  [->,thin,gray](1.9,0)..controls ++(down:4mm) and ++(left:5mm).. (2.75,-1.45) 
..controls ++(right:4mm) and ++(down:5mm)..(3.6,0);
\draw  [->,thin,gray] (2.75,-2.6) --  (2.75,0);
\draw  [->,thin,gray] (2.75,0) --  (2.75,2.6);
\node[below] at (4.5,0.5) {$=$} ;
 \draw [->,red,thick] (6,-3) -- (6,3)  ;
 \draw [->,blue,thick] (7,-3) -- (7,3);
  \filldraw [green] (5.5,0) circle (2pt) ;\node[above] at (5.5,0) {$b$};
  \filldraw (6.5,0) circle (2pt) ;\node[above] at (6.5,0) {$a$};
 \filldraw  [green](7.5,0) circle (2pt) ;\node[above] at (7.5,0) {$c$};
\end{tikzpicture} 
\end{center}
\caption{Identities related to Reidemeister moves I and II. On the left
there is a summation over the internal spin $a$, so that $c'=c$. On the
right there is a summation over internal spin $d$, so that $a'=a$.}
\label{fig4}\end{figure}
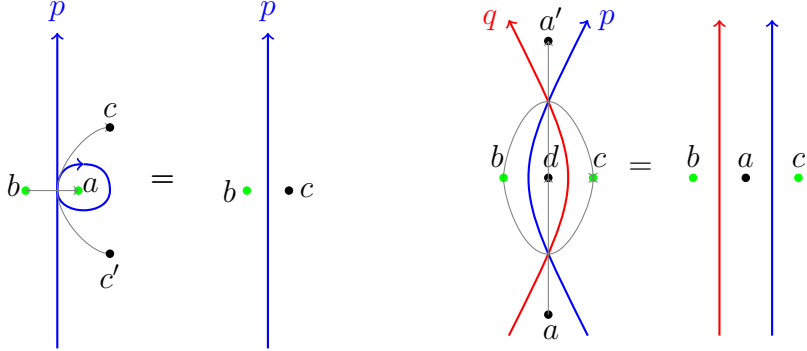
\ba
&&W_{pp}(n)=\Big(\frac{\mu_p}{\mu_p}\Big)^{\!\!n}\prod_{j=1}^n
\frac{y_p-x_p\omega^j}{y_p-x_p\omega^j},\quad\hbox{or}\quad
{W_{pp}(a-b)=1},\\
&&\overline W_{pp}(n)=\big({\mu_p\mu_p}\big)^{\!n}\!\prod_{j=1}^n
\frac{\omega x_p\!-\!x_p\omega^j}{y_p\!-\!y_p\omega^j} ,\quad\hbox{or}\quad
{\overline W_{pp}(a-b)=\delta_{a,b}}.
\ea
corresponds to Reidemeister move I and
 \ba
W_{pq}(n)\!=\!\Big(\frac{\mu_p}{\mu_q}\Big)^{\!\!n}\prod_{j=1}^n
\frac{y_q-x_p\omega^j}{y_p-x_q\omega^j}=W^{-1}_{qp}(n),\\
\hbox{or}\quad{W_{pq}(b-c)W_{qp}(b-c)=1},\nonumber\\
{\sum_{d=0}^{N-1}\overline W_{pq}(a-d)\overline W_{qp}(d-a')=r_{pq}\delta_{a,a'}},
\ea
with some factor $r_{pq}$, corresponds to Reidemeister move II. These
are depicted in figure \ref{fig4}.

\begin{figure}
\begin{center}
\begin{tikzpicture}[scale=0.48]
\draw  [->,blue,thick](3,-5)  -- (3,5) node[left]  {$p$} ;
\draw  [->,red,thick] (5,5) --(-2,-2)  node[left] {$q$} ;
\draw  [->,green,thick] (5,-5) -- (-2,2) node[left] {$r$} ;
 \filldraw  (-1.5,0) circle (3pt) ;\node[left] at (-1.5,0) {$b$};
 \filldraw  (1.8,0) circle (3pt) ;\node[right] at (1.8,0) {$d$};
  \filldraw (3.6,4.5) circle (3pt) ;\node[above] at (3.6,4.5) {$c$};
   \filldraw (3.6,-4.5) circle (3pt) ;\node[below] at (3.8,-4.5) {$a$};
     \draw  [->,thin,gray](3.6,-4.5)  --  (2.7,-2.25);\draw  [thin,gray] (2.7,-2.25) -- (1.8,0);
       \draw  [->,thin,gray] (1.8,0) -- (2.7,2.25) ;\draw  [thin,gray] (2.7,2.25)  -- (3.6,4.5);
          \draw  [->,thin,gray] (1.8,0) --  (-1.0,0);\draw  [thin,gray] (-1.0,0) -- (-1.5,0);
\node[below] at (7,0.5) {$=$} ;
 \draw  [->,blue,thick](12,-5)  -- (12,5) node[right]  {$p$} ;
\draw  [->,red,thick] (16,2) --(9,-5)  node[left] {$q$} ;
\draw  [->,green,thick] (16,-2) -- (9,5) node[left] {$r$} ;
  \filldraw  (10,0) circle (3pt) ;\node[left] at (10,0) {$b$};
  \filldraw (14,4) circle (3pt) ;\node[right] at (14,4) {$c$};
   \filldraw (14,-4) circle (3pt) ;\node[right] at (14,-4) {$a$};
 \draw  [->,thin,gray] (14,-4) --   (14,2); \draw  [thin,gray] (14,2) --   (14,4);
\draw  [->,thin,gray](14,-4)  --  (11,-1); \draw  [thin,gray] (11,-1) --   (10,0);
\draw  [->,thin,gray] (10,0) --(13,3) ; \draw  [thin,gray] (13,3) --   (14,4);
\end{tikzpicture}
\end{center}
\caption{Spin-model Yang--Baxter (or Star-Triangle) Equation generalizing
Reidemeister move III.}
\label{fig5}\end{figure}
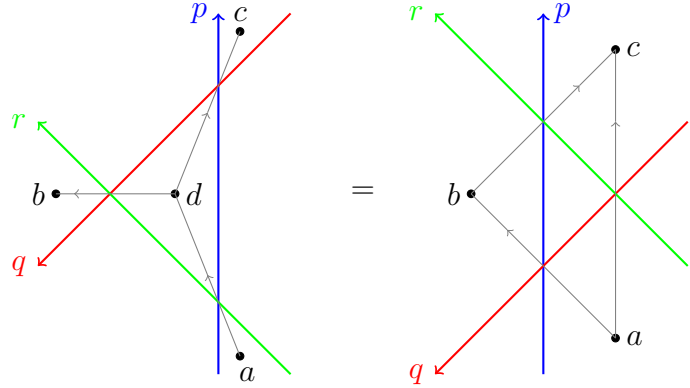

Finally there is the Star-Triangle Relation (Yang--Baxter relation for
the chiral Potts model), see also figure \ref{fig5},
\ba
&&\sum^{N}_{d=1}\,{\overline W}_{pr}(a-d)\,W_{pq}(d-c)\,{\overline  W}_{rq}(d-b)
\label{StarTriangle}\\
&&\qquad\qquad=R_{pqr}\,\overline W_{pq}(a-b)\,{ W}_{pr}(b-c)\,W_{rq}(a-c),\nonumber
\ea
corresponding to Reidemeister move III.

As is well known, repeated application of the Star-Triangle Relation, see
figure \ref{fig6}, implies that transfer matrices commute. Defining
\begin{figure}[ht]
\begin{center}
\begin{tikzpicture}[scale=0.4]
 \foreach \x in {2,...,9}
\draw  [->,blue](2*\x,80pt)  -- (2*\x,160pt) node[above]  {$p$} ;
 \draw [->,red](19,150pt) --(3.5,150pt)..controls ++(left:8mm) and ++(up:8mm)..(1,80pt ) node[left] {$q$};
 \draw [->,green](19,90pt) --(3.5,90pt)..controls ++(left:8mm) and ++(down:8mm)..(1,160pt ) node[left] {$r$};
 \node[right] at (19,150pt) {$\hat T_q$} ;\node[right] at (19,90pt) {$ T_r$} ;
 \end{tikzpicture}
\end{center}
\end{figure}
\begin{figure}[ht]
\begin{center}
$=$
\end{center}
\end{figure}
\begin{figure}[ht]
\begin{center}
\begin{tikzpicture}[scale=0.4]
 \foreach \x in {2,...,9}
\draw  [->,blue](2*\x,-40pt)  -- (2*\x,40pt) node[above]  {$p$} ;
   \draw [->,green] (20,-40pt)..controls++(up:8mm) and ++(right:8mm)..(18.5,30pt)-- (3,30pt) node[left] {$r$};
 \draw [->,red](20,40pt) ..controls++(down:8mm) and ++(right:8mm)..(18.5,-30pt)-- (3,-30pt) node[left] {$q$};
  \node[left] at (2,30pt) {$\hat T_r$} ;\node[left] at (2,-30pt) {$ T_q$} ;
 \end{tikzpicture}
\end{center}
\caption{Commuting Transfer Matrices: Shown here is the result of
repeated application of the Star-Triangle Equation. Applying an additional $qr$
weight to the right before closing the $q$ and $r$ rapidity lines, we can apply
the inversion relation and prove that the transfer matrices commute.}
\label{fig6}\end{figure}
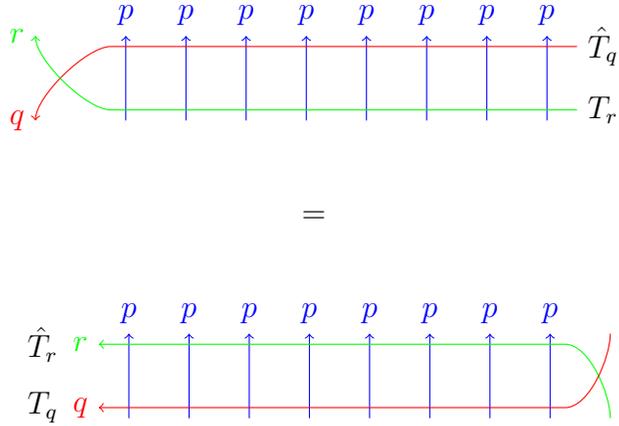
\ba
&&({T}_q)_{\sigma,\sigma'}=\prod_{j=1}^L\,W_{pq}(\sigma_j,\sigma'_j)
{\overline W}_{pq}(\sigma_{j+1},\sigma'_j),\\
&&(\hat{ T}_r)_{\sigma,\sigma'}=\prod_{j=1}^L\,\overline W_{pr}(\sigma_j,\sigma'_j)
{ W}_{pr}(\sigma_j,\sigma'_{j+1}),\nonumber
\ea
with periodic boundary conditions $\sigma_{L+1}=\sigma_1$ and
$\sigma'_{L+1}=\sigma'_1$, we can prove \cite{Baxterbook}
\be
{{T}_q\hat{ T}_r\propto{T}_r\hat{ T}_q.}
\ee
More precisely, combining the chiral Potts weights in two successive diagonal rows,
as indicated in figure \ref{fig6}, and summing over all spins in the middle row,
we get the two products of the two transfer matrices, after closing the horizontal
$p$ and $q$ rapidity lines. Inserting an inversion relation, we can then repeatedly
apply (\ref{StarTriangle}).

\section{Relation with Hypergeometric Series}
The Boltzmann weights $W_{pq}(n)$ and $\overline W_{pq}(n)$ for edges
between spins $a$ and $b$, $n=a-b$, in (\ref{Bweights}) can be rewritten
in terms of $\omega$-Pochhammer symbols as
\ba
W_{pq}(n)\!=\!\Big(\frac{\mu_p}{\mu_q}\Big)^{\!\!n}\prod_{j=1}^n
\frac{y_q-x_p\omega^j}{y_p-x_q\omega^j}=\gamma_{pq}^n
\frac{(\alpha_{pq},\omega)_n}{(\beta_{pq},\omega)_n},\\
  \overline W_{pq}(n)\!=\!\big({\mu_p\mu_q}\big)^{\!n}\!\prod_{j=1}^n
\frac{\omega x_p\!-\!x_q\omega^j}{ y_q\!-\!y_p\omega^j} 
={\bar\gamma}_{pq}^n\frac{(\bar\alpha_{pq},\omega)_n}{(\bar\beta_{pq},\omega)_n},
\nonumber\ea
where
\ba 
&&{(x,\omega)_n=\prod_{\ell=0}^{n-1}(1-x\omega^\ell),}\quad
\omega={\rm e}^{2\pi \sqrt{-1}/N},\\
&&\gamma_{pq}={\mu_p y_q}/{\mu_q y_p},\qquad
\alpha_{pq}={\omega x_q}/{y_p},\quad
\beta_{pq}={\omega x_p}/{ y_q},\\
&&{\bar\gamma}_{pq}={\omega\mu_p\mu_q x_p}/{ y_q},\quad
\,\bar\alpha_{pq}={x_q}/{ x_p},\quad\bar\beta_{pq}={\omega y_p}/{ y_q},
\nonumber\ea
It is obvious that the star-triangle relation is related to identities of
basic hypergeometric series. 

\subsection{Basic Hypergeometric Series}
Define
\be
\hypg{a_1,a_2,\cdots,a_{p+1}}
{\hphantom{aa}b_1,b_2\cdots,b_p}{z}=
\sum_{l=0}^{\infty}\frac{(a_1;q)_l(a_2;q)_l\cdots(a_{p};q)_l(a_{p+1};q)_l}
{(b_1;q)_l(b_2;q)_l\cdots(b_{p};q)_l(q;q)_l}\,z^{l},
\label{defbhs}\ee
with the $q$-Pochhammer symbol 
\be
(x;q)_l=\prod_{j=0}^{l-1}(1-xq^{j}), \quad |q|<1,
\ee
generalizing the usual Pochhammer symbol
\be
(a)_l=a(a+1)\cdots(a+l-1).
\ee
The series defined in (\ref{defbhs}) is then the $q$-deformation of the generalized hypergeometric series
\be
\hypf{p+1}{p}{a_1,a_2,\cdots,a_{p+1}}
{\hphantom{aa}b_1,b_2\cdots,b_p}{z}=
\sum_{l=0}^{\infty}\frac{(a_1)_l(a_2)_l\cdots(a_{p})_l(a_{p+1})_l}
{(b_1)_l(b_2)_l\cdots(b_{p})_l\,l!}\,z^{l}.
\label{defgF}\ee
For the root-of-unity case with $\omega^N=1$, we find
\be 
\prod_{j=0}^{N-1}(1-x\omega^{j})=(1-x^N)\to(\omega;\omega)_N=0.
\ee
Thus in the limit $q\to\omega$, the summand in the series in (\ref{defbhs})
is divergent. To have a well-defined summand, we have to let $a_{p+1}=q^{1-N}$
to make it a terminating series, that is
\be
\hypg{\omega,\alpha_1,\cdots,\alpha_{p}}
{\hphantom{aa}\beta_1,\cdots,\beta_p}{z}=
\sum_{l=0}^{N-1}
{\frac{(\alpha_1;\omega)_l\cdots(\alpha_{p};\omega)_l}
{(\beta_1;\omega)_l\cdots(\beta_{p};\omega)_l}}\,z^{l}.
\label{defbhsc}\ee

Many well-known theorems in basic hypergeometric series may not hold
in the root-of-unity case, and need to be modified, such as the transformation
formula

\noindent
{\bf{\it Theorem} 10.2.1} \cite{AAR}
 \be
\sum_{n=0}^{\infty}\frac{(a;q)_n} {(q;q)_n} x^n=
\frac{(ax;q)_{\infty} }{(x;q)_{\infty}},\quad \hbox{for}\quad |x|<1
\label{10.2.1}\ee
which is the $q$-analog of the binomial theorem
\be
\sum_{n=0}^{\infty}\frac{(a)_n} {n!} x^n=
(1-x)^{-a}\quad \hbox{for}\quad |x|<1.
\ee
Only for the case with $a=q^{-\alpha}$, can (\ref{10.2.1}) be extended
to the root-of-unity case as
\be
\sum_{n=0}^{\alpha}\frac{(\omega^{-\alpha};\omega)_n} {(\omega;\omega)_n} x^n
=(\omega^{-\alpha}x;\omega)_{\alpha}
=\sum_{n=0}^{\alpha}\left[{\hidefrac{\alpha}{n}}\right]
(-x)^n\omega^{{\scriptstyle{\frac 12}} n(n-1)-n\alpha}.
\ee
This well-known formula is due to Rothe, according to \cite{AAR}.
 
Similarly, the $q$-analog of Euler's formula 
\be
\hypf{2}{1}{a,b}{\hphantom{a}c}{x}=
(1\!-\!x)^{c-a-b}\hypf{2}{1}{c\!-\!a,c\!-\!b}{\hphantom{aa}c}{x}
\ee
{ {\bf {\it Theorem} 10.10.1} \cite{AAR} }
\be
\hypp{2}{1}{a,b}{\hphantom{a}c}{x}=\frac{(abx/c;q)_\infty} {(x;q)_{\infty}}
\hypp{2}{1}{c/a,c/b}{\hphantom{aa}c}{\frac{abx}c},
\ee
does not hold for $q\to\omega$. However, for some particular values
of $a$, $b$ and $c$, we find
\ba
&&\label{10.10.1c}\\
&&
\hypp{2}{1}{\omega^{\alpha},\omega^{\beta}}{\hphantom{\omega}\omega^{\gamma}}{t}
=
\big(\omega^{\alpha+\beta-\gamma}t;\omega\big)_{\!N-\alpha-\beta+\gamma}\,\,
\hypp{2}{1}{\omega^{\gamma-\alpha},\omega^{\gamma-\beta}}
{\hphantom{\omega,,}\omega^{\gamma}}{\omega^{\alpha+\beta-\gamma}t}.\nonumber
\ea
The proof was rather non-trivial, as will be outlined later. This shows that the
summation formulae of the basic hypergeometric series cannot be extended
to the root-of-unity case, unless further restrictions are imposed.

\subsection{Cyclic Hypergeometric Series}
If we impose the periodicity requirement for the finite sum in (\ref{defbhsc}) to be
\be
z^N=\prod_{j=1}^p\gamma_j^N,\qquad
{\gamma_j}^N=\frac{1-\beta_j^N}{1-\alpha_j^N},
\ee
we obtain a ``cyclic hypergeometric function" with summand
periodic mod $N$. 

The Fourier transform of the chiral Potts weight is a cyclic ${}_2\myphi_1$, i.e.
\ba
W^{({\rm f})}(k)=\sum_{n=0}^{N-1}\omega^{nk}\,W(n)=
\hypp{2}{1}{\omega,\alpha}{\hphantom{\omega}\beta}{\gamma\,\omega^k},
\ea
where
\be
W(n)=\gamma^{n}\,\frac{(\alpha;\omega)_n}{(\beta;\omega)_n},\quad W(N+n)=W(n),
\quad\gamma^N=\frac{1-\beta^N}{1-\alpha^N}.
\ee
It is summable as shown in \cite{BBP,APmf}, namely
\be
\hypp2{1}{ \omega,x}{\hphantom{\omega} y}{z}=
 \frac{\omega^d N^{\frac 1 2}}{\Phi_{ 0} \Delta\!(y)^{N\! -\!1}}
\frac{p(y)p(\omega x/y)p( z)}{ p( x) p(\omega xz/y)},
\label{phi21c}\ee
where
\ba
{z^N(1-x^N)=(1-y^N)},\quad p(x) = \prod_{j=1}^{N-1}(1-\omega^{j}x)^{j/N},\\
\Delta\!(x)=(1-x^N)^{1/N},
\quad\Phi_0=e^{i\pi(N-1)(N-2)/12N}.\nonumber
\ea
However, the periodic restriction make the Riemann sheet structure very complicated.
It was also shown in \cite{APmf} the following relations hold
\be
\hypp21{\omega,x}{\hphantom{\omega}y}{z}
{\hypp21{\omega,y/xz}{\hphantom{\omega}\omega/z}{x}}=N.
\label{wff}
\ee
and
\ba
&&\hypp2{1}{\omega,x\omega^m}{\hphantom{\omega}y\omega^n}{z\omega^k}
\label{rec4}\\
&&\quad=\hypp2{1}{\omega,x}{\hphantom{\omega}y}{z}
\left(\frac{\omega}{ y}\right)^{\! k} \!(z\omega^{k})^{-n}
\frac{(y;\omega)_n (z;\omega)_k
(\omega x/y;\omega)_{m-n}}{(x;\omega)_m(\omega xz/y;\omega)_{m-n+k}}.\nonumber
\ea

\subsection{ Cyclic Hypergeometric  ${}_{3}\myphi_{2}$}
It is found in \cite{APhyp} that the cyclic hypergeometric  ${}_{3}\myphi_{2}$ satisfies the transformation formula\footnote{It should be noted that Sergeev, Mangazeev, and
Stroganov \cite{SMS} derived similar identities, using an upside-down version of
the $q$-Pochhammer symbol.}
\be
\hypp3{2}{\omega,x_1,x_2}{\hphantom{\omega}y_1,y_2}{z}=A\,\,\,
\hypp3{2}{\omega,\hphantom{a}{z/z_1},\hphantom{a}{y_1/x_1z_1}}
{\hphantom{aa,}{\omega/z_1},{\omega x_2 z/y_2z_1}}
{\frac{\omega x_1}{y_2}},
\label{phi32c}\ee
where the periodic restriction is
\be 
{z^N(1-x_1^N)(1-x_2^N)=(1-y_1^N)(1-y_2^N)},
\label{periodic}\ee
and the constant
\be
A=N^{-1}
\hypp2{1}{\omega,x_1}{\hphantom{,\omega}y_1}{z_1}
\hypp2{1}{\omega,x_2}{\hphantom{,\omega}y_2}{\frac{z}{z_1}},\quad 
z_1^N=\frac{(1-y_1^N)}{(1-x_1^N)}.
\ee
If $z=\omega$, we find ${}_{3}\myphi_{2}$ on the right-hand side of (\ref{phi32c}) becomes
${}_{2}\myphi_{1}$, so that it is a product of three cyclic ${}_{2}\myphi_{1}$, which are summable. Therefore we find ${}_{3}\myphi_{2}$ is also summable at $z=\omega$. Now (\ref{periodic}) becomes
\be
(1-x_1^N)(1-x_2^N)=(1-y_1^N)(1-y_2^N),
\ee
which gives rise to a very complicated Riemann sheet structure.

We next outline the proof of (\ref{10.10.1c}). Consider the series
in (\ref{phi32c}) with $x_1=x$, 
$x_2=\omega^{\gamma-\beta}y$, $y_1=\omega^{\alpha}x$,
$y_2=\omega y$ and $z=\omega^\beta$.
In this case, it becomes
\be
\hypp3{2}{\omega,x,\omega^{\gamma-\beta}y}
{\hphantom{\omega}\omega^\alpha x,\omega y}{\omega^\beta}=
{\bar B}\,\hypp2{1}{\omega^\beta,\omega^\alpha}
{\hphantom{\omega,}\omega^\gamma}{\frac x y},
\ee
where
\be
{\bar B}=N^{-1}
\hypp2{1}{\omega,x}{\hphantom{\omega}\omega^\alpha x}{1}
\hypp2{1}{\omega,\omega^{\gamma-\beta}y}{\hphantom{\omega}\omega y}{\omega^\beta}.
\ee
If we interchange $x_1$ and $x_2$, and then use (\ref{phi32c}) twice, we obtain
\ba
\hypp3{2}{\omega,\omega^{\gamma-\beta}y,x}{\hphantom{\omega,}\omega^\alpha x,
\hphantom{a}\omega y}{\omega^\beta}&=&
A\,\hypp3{2}{\omega,\omega^{\beta}z,\omega^{\alpha+\beta-\gamma}zx/y}
{\omega z,\hphantom{\omega}\omega^\beta z x/y}{\omega^{\gamma-\beta}}\\
&=&A B\,
\hypp2{1}{\omega^{\gamma-\beta},\omega^{\gamma-\alpha}}{\hphantom{\omega,}\omega^\gamma}
{\omega^{\alpha+\beta-\gamma}\frac x y},\nonumber
\ea
where the constants are
\ba
&&A=N^{-1}
\hypp2{1}{\omega,\omega^{\gamma-\beta}y}{\omega^\alpha x}{1/z}
\hypp2{1}{\omega,x}{\hphantom{\omega}\omega y}{\omega^\beta z},\\
&&B=N^{-1}
\hypp2{1}{\omega,\omega^{\alpha+\beta-\gamma}zx/y}{\omega^{\beta}zx/y}{1}
\hypp2{1}{\omega,\omega^{\beta}z}{\hphantom{\omega}\omega z}{\omega^{\gamma-\beta}}.
\nonumber\ea
Now we may use (\ref{rec4}), (\ref{wff}) and (\ref{phi21c}) to find
\be
AB/{\bar B}=(\omega^{\alpha+\beta-\gamma}x/y;\omega)_{N-\alpha-\beta+\gamma}.
\ee

\subsection{ Cyclic Hypergeometric  ${}_{4}\myphi_{3}$} 
Clearly, the star-triangle relation (\ref{StarTriangle}) gives a summation formula for
${}_{4}\myphi_{3}$. To convert the left-hand side of the star-triangle equation
into ${}_{4}\myphi_{3}$, we must rewrite the Pochhammer symbols in the weights
in terms of the rapidities. We find
\ba
&&\alpha_1=\omega^{c-a-1}y_r/y_p,\quad \alpha_2=\omega x_p/y_q,
\quad \alpha_3=\omega^{c-b} x_q/x_r,\\
&&\beta_1=\omega^{c-a}x_p/x_r,\quad \beta_2=\omega  x_q/y_p,
\quad \beta_3=\omega^{c-b+1}y_r/y_q,\nonumber\\
&&\gamma_1=y_p/\mu_p\mu_r x_r,\quad \gamma_2= \mu_p y_q/\mu_q y_p,
\quad \gamma_3=\omega\mu_q\mu_r x_r/y_q.\nonumber
\ea
This gives
\be
\omega^2\alpha_1\alpha_2\alpha_{3}=\beta_1\beta_2\beta_3,\quad
\gamma_1\gamma_2\gamma_3=\omega.
\ee
It is known that when the
{well-balanced condition}
\be
q a_1 a_2\cdots a_{p+1}=b_1 b_2\cdots b_p,\quad z=q,
\ee
 is satisfied, there exist many summation formulae for the basic hypergeometric series.
The most well-known summation formula is
\be
\hypp{3}{2}{a,b,\hphantom{\omega}q^{-n}\hphantom{\omega}}{\hphantom{a,}c,q^{1-n}ab/c}{q}=\frac{(c/a;q)_n(c/b;q)_n} {(c;q)_n(c/ab;q)_n},
\ee
which is the $q$-analog of the Pfaff-Saalsch\"utz formula
\be
\hypf{3}{2}{a,b,\!-\!n}{c,1\!+\!a\!+\!b\!-\!n\!-\!c}{1}
=\frac{(c-a)_n(c-b)_n} {(c)_n(c-a-b)_n}.
\ee

\subsection{\boldmath{$N\to\infty$} Limits}
By taking the $N\to \infty$ limit \cite{APinf}, the star-triangle relation becomes the summation formula for double sided series 
\be
\sum_{n=-\infty}^{\infty}
\frac{\Gamma(x_1+n)\Gamma(x_2+n)\Gamma(x_3+n)}{\Gamma(y_1+n)
\Gamma(y_2+n)\Gamma(y_3+n)}=
\frac{G(x_1,x_2,x_3\vert y_1,y_2,y_3)}
{\prod_{i=1}^{3}\prod_{j=1}^{3}\Gamma(y_i-x_j)},
\ee
if both the well-balanced condition and the
periodicity condition hold, i.e.
\ba
&&x_1+x_2+x_3+2=y_1+y_2+y_3,\\
&&\sin\pi x_1\,\sin\pi x_2\,\sin\pi x_3=
\sin\pi y_1\,\sin\pi y_2\,\sin\pi y_3,
\nonumber\ea
where
\ba
&&G(x_1,x_2,x_3\vert y_1,y_2,y_3)\\
&&\qquad\equiv
\prod_{j=2}^3\Gamma(x_j)\Gamma(1-x_j)
\prod_{i=1}^3\Gamma(y_i-x_1)\Gamma(1-y_i+x_1),\nonumber
\ea
which is invariant under
\begin{itemize}
\item[1$^\circ$] {Permutations of $x_1,x_2,x_3$ and $y_1,y_2,y_3$,}
\item[2$^\circ$]  {Reflections $x_j\mapsto1-y_j$,
$y_j\mapsto1-x_j$ simultaneously,}
\item[3$^\circ$]  {Translations $x_j\mapsto x_j+M$,
$y_j\mapsto y_j+M$ for $j=1,2$ or $3$.}
\end{itemize}

We note that there are also other $N\to\infty$ limits with spins taken from a finite
or infinite continous interval \cite{APinf}.

\section{Final remark}

At the Qinhuangdao meeting we also discussed the superintegrable subcase 
of the integrable chiral Potts model, which has an additional underlying Onsager
loop group structure, and we discussed how the spectrum is then dominated by an
affine quantum group at an $N$-th root of unity, where $N$ can be odd or even.
Even though the theory of this quantum group is much better understood for odd 
roots of unity, we presented new approaches to establish the higher quantum
Serre relations also for the even root-of-unity case. We shall not go in more detail
here as a more complete account has been presented at the meeting in honor
of Baxter's 75th birthday \cite{JPBax,APserre}.


\end{document}